\documentclass[12pt]{article}
\usepackage[cp1251]{inputenc}
\usepackage[english]{babel}
\usepackage{graphicx,graphics}
 0
\begin{document}
\title{Zero-Range Potentials in Multi-Channel Diatomic Molecule Scattering}
\author{  S.B. Leble
\small \\ Theoretical Physics and Mathematical Methods Department,
\small\\ Technical University of Gdansk, ul, Narutowicza 11/12, Gdansk,  Poland,
\small \\ leble@mifgate.pg.gda.pl \\  \\[2ex]
S. Yalunin \small\\ Theoretical Physics Department, \small\\
Kaliningrad State University, A. Nevsky st. 14, Kaliningrad,
Russia}

\maketitle

\renewcommand{\abstractname}{\small Abstract}
 \begin{abstract}
The method of zero-range potentials is generalized to account for the
molecular electron excitation process. It is made by a matrix
formulation in which a state vector components are associated with
a scattering channel. The multi-center target is considered and
the model is applied to the example of $e + H_2$ low energy
scattering. The results of evaluation of cross-sections are
compared with ones of the MCF and SMC methods.
\end{abstract}

\thispagestyle{empty}

\section {Introduction.}
 The ideas of zero range potential
(ZRP) approach were recently developed to widen limits of the
traditional treatment  \cite{DO1975}, the author accounts higher
momentum partial amplitudes \cite{Bal} for a multi-center problem.
Historically, the ZRP notion was introduced, perhaps, in \cite{F}
to model a potential action sustaining some parameters values. The
simplification of the theory is such that a problem of
differential equation goes down to some algebraic one \cite{B}.
There is the book \cite{DO1975} and the review $\cite{La1980}$ in
which the theory is comprehensively described.

The target of this paper relates to the other limitation of the
theory connected  with multi-channel character of real scattering
phenomenon. Following the ideas of \cite{DO} we introduce  a
vector formulation of the scattering problem with the components
that stands for different possibilities (channels) of the process
to be described. The scattering amplitude hence is a matrix
connecting asymptotic states. In the Sec. 1 the formalism is
introduced for a multi-center problem.

As an important example we consider applications to a diatomic
molecule. In the framework of the generalization to be introduced
we revisit results of the papers \cite{DR1970}, \cite{DYu1977}
starting from adiabatic nuclei problem (Sec.2). The rotations of a
molecule are included similarly to the mentioned approach
\cite{DYu1977} but with some difference in the formalism.
Oscillations are described by means of the Morse potential with
eigen functions - vibrational harmonics that are proportional to
Laguerre polynomials (Appendix). The results of differential  and
integral cross-sections evaluation are compared with the direct
numerical solution by some standard molecular orbits model
\cite{Lim1988} (Sec.3).

\section{The matrix zero-range potentials}
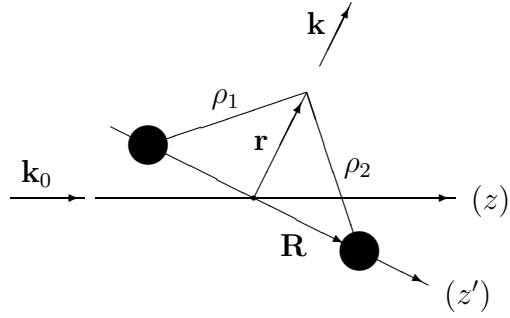
\begin{figure*}
\setlength{\unitlength}{2pt}
\center{\begin{picture}(90,60)(0,8)
\put(15,25){\line(68,0){68}} \put(78,25){\vector(4,0){4}} \put(86,23){$({z})$}
\put(18,38.5){\line(2,-1){60}} \put(73,11){\vector(2,-1){4}} \put(81,5){$({z}')$}
\put(45,25){\circle*{1}}
\put(25,35){\circle*{8}}
\put(65,15){\circle*{8}}
\put(-1,25){\line(14,0){14}} \put(8,25){\vector(4,0){4}} \put(1,28){${\bf k}_0$}
\put(45,25){\line(1,2){10}} \put(50,35){\vector(1,2){4}} \put(45,34){${\bf r}$}
\put(25,35){\line(3,1){30}} \put(37,43){$\rho_{1}$}
\put(65,15){\line(-1,3){10}} \put(62,30){$\rho_{2}$}
\put(57.5,50){\line(1,2){6}} \put(59,53){\vector(1,2){4}} \put(55,56){${\bf k}$}
\put(58,18.5){\vector(2,-1){4}} \put(50,14){$\bf R$}
\end{picture}}
\caption{Two-center model. The ZRP are represented by full circles.} \label{f:1}
\end{figure*}

\setcounter{equation}{0} As it is known, if we diminish the range
and at the same time increase the depth of a pertinent spherically
symmetrical potential then
%the potential can be concentrated at
%the center of symmetry. At that one
the only characteristic parameter of a potential may remain fixed
(usually it's a scattering length or a position of the
bound-state energy level), others characteristic are lost. This
procedure allow to replace the potential action by the boundary
condition on the wave function $\psi({\bf r})$ at the potential
center (see Ref. $\cite{DO1975,D1978}$)
$$
\left[ \frac{\partial}{\partial \rho_j}\, \ln (\rho_j\, \psi ) \right]_{\,\rho_j=0}=
- \, \alpha_j, \hspace{15mm} \rho_j=|{\bf r}\, {\scriptstyle -}\, {\bf R}_j|,
$$
where $\alpha_j$ - inverse scattering length, ${\bf R}_j$ -
position vector of the potential center $j$. It's necessary to
remark that ZRP can be introduced in another ways. For example, it
can be defined by the following equality (see Ref. $\cite{DO},
\cite{AlGa1988}$)
$$
V_j=\frac{2\pi}{\alpha_j}\, \delta(\mbox{\boldmath$\rho$}_j)\,
\frac{\partial}{\partial \rho_j}\,
\rho_j,
$$
where $\delta(\mbox{\boldmath$\rho$}_j)$ - Dirac function in the
three-dimensional space. In order to adapt ZRP
model for multichannel scattering we will replace the parameters
$\alpha_j$ by matrices $A_j$ and a wave function $\psi({\bf r})$
by the vector function $\Psi({\bf r})$
$$
\Psi({\bf r})=
\left(
\begin{array}{c}
\psi_{0}({\bf r}) \\ \ldots \\ \psi_{N}({\bf r})
\end{array}
\right).
$$
The components $\psi_{n}({\bf r})$ of the vector function
$\Psi({\bf r})$ have the form (see Ref. $\cite{DO1975}$)
\begin{equation} \label{PSI}
\psi_{n}({\bf r},{\bf R})=e^{\displaystyle i({\bf k}_0,{\bf
r})}\ \delta_{n 0} + \sum_{j}\, (C_j)_{n}\, \frac{e^{\displaystyle i k_{n}
\rho_{j}}}{\rho_{j}}.
\end{equation}
Here $C_{j}$ are some constant vectors, $k_{n}$ - electron's
momenta in the channel $n$, ${\bf k}_{0}$ - momentum of the
incoming electron. The representation $(\ref{PSI})$ indicates the
component $\psi_{n}({\bf r})$ describes the electron scattering in
the channel $n$. The boundary conditions for components
$\psi_{n}({\bf r})$ have the form
\begin{equation}\label{BC1}
\hspace{20mm}\left[
\frac{\partial}{\partial \rho_j}\, \rho_j\, \psi_{n} \right]_{\rho_j=0}
= - \left[ \sum_{n'=1}^{N}\, \rho_j\, (A_j)_{nn'} \psi_{n'} \right]_{\rho_j=0}.
\end{equation}
This equation is basic in the ZRP theory. General properties of the
matrices $A_j$ are determined by symmetry of the quantum states.
It means the matrices of different centers are
linked by symmetry transformation. For example, in the case of
only two centers the matrix $A_2$ is a transform of $A_1$.
\paragraph{\hspace{5mm} \it The diatomic, homogeneous molecules at the
N-$\Sigma$-state  approximation:} Let us consider the problem of
electron-impact excitation of a diatomic, homogeneous molecule in
the $N$-state model.
%Let us consider
The coordinate systems will have the common origin at the center
of mass of a molecule. Then ${\bf R}_1=-{\bf R}$ and ${\bf
R}_2={\bf R}$, here $2R$ is the internuclear distance (see Fig.
$\ref{f:1}$). The space-fixed axis $oz$ of the first frame is
chosen along initial momentum ${\bf k}_0$ (the so-called
LAB-frame, see the review $\cite{La1980}$), whereas axis $oz'$ of
the second system we direct along a symmetry axis of the molecule
(i.e. along ${\bf R}$, such coordinate system is known as
BODY-frame).

Though atoms in the molecule are identical the off-diagonal
elements of the atom potentials $V_{1,2}$ (and therefore matrices
$A_{1,2}$) can be different in sign because the parities of the
molecular states $|n \rangle$ can be different. Suppose hence
\begin{equation}
A_1=  \hbox{diag}(\alpha_{0},\ldots,\alpha_{N}) + A,
\end{equation}
if $A$ is some Hermitian matrix with zero in-diagonal elements.
Let us consider the diagonal matrix
$\sigma_{\eta}=\hbox{diag}(\eta_0,\ldots,\eta_N)$, where
$\eta_{n}$ are parities of molecular states $|n \rangle$. Then the
matrix $A_2$ is given by
\begin{equation}
A_2 = \sigma_{\eta}\, A_1\, \sigma_{\eta} =
\hbox{diag}(\alpha_{0},\ldots,\alpha_{N}) + \sigma_{\eta}\,
A\, \sigma_{\eta},
\end{equation}
where $\alpha_{n}$ are some parameters of the molecular states $|n\rangle$,
$A$ - matrix of the coupling channel parameters.
\setlength{\unitlength}{1pt}

If the nuclei are held space-fixed then the components $\psi_{n}({\bf r})$
of the vector function $\Psi({\bf r})$ have the form
\begin{equation} \label{PSI2}
\psi_{n}({\bf r})=e^{\displaystyle i({\bf k}_0,{\bf
r})}\ \delta_{n 0} + (C_1)_{n}\, \frac{e^{\displaystyle i k_{n}
\rho_{1}}}{\rho_{1}} + (C_2)_{n}\, \frac{e^{\displaystyle i k_{n}
\rho_{2}}}{\rho_{2}}, \ \ \ \ \ \rho_{1,2}=|{\bf r}\, {\scriptstyle \pm}\, {\bf R}|.
\end{equation}
Therefore transition amplitudes in fixed-nuclei approximation can
be constructed by the expressions
\begin{equation} \label{alplitudes}
f_{n}({\bf k},{\bf k}_0,{\bf R})=(C_1)_{n}\, e^{\, \displaystyle i({\bf k},{\bf
R})} + (C_2)_{n}\, e^{\, -\displaystyle i({\bf k},{\bf R})},\ \ \ \ \ \ \
{\bf k}=k_{n}\, \frac{{\bf r}}{r},
\end{equation}
where ${\bf k}$ - outgoing electron momentum.

It is necessary to note that other notations are more convenient
\begin{equation} \label{C-S}
(C_1)_n = S^{(+)}_n + S^{(-)}_n,
\hspace{10mm}
(C_2)_n = \eta_0 \eta_n \left(
S^{(+)}_n - S^{(-)}_n \right).
\end{equation}
Using the boundary conditions $(\ref{BC1})$ we
 obtain the vectors $S^{(+)}$ and $S^{(-)}$. To introduce
the matrices
$$
\Lambda^{(\pm)}=\hbox{diag}(\theta^{(\pm)}_0,\ldots,\theta^{(\pm)}_N)
+A,
$$
where we used the following notation
$$
\theta^{(\pm)}_{n}= \alpha_n+ik_n \pm \eta_0 \eta_n
\frac{e^{\displaystyle 2ik_n R}}{2R},
$$
one obtains for $S^{(+)}$ and $S^{(-)}$ the expressions
\begin{equation} \label{L-eq}
S^{(+)}_{n} = - \cos({\bf k}_0,{\bf R})\, (\Lambda^{(+)-1})_{n0},\hspace{10mm}
S^{(-)}_{n} = i \sin({\bf k}_0,{\bf R})\, (\Lambda^{(-)-1})_{n0}.
\end{equation}
Using Eqs. $(\ref{alplitudes})$, $(\ref{C-S})$ and $(\ref{L-eq})$
we arrive to the observation that if $\eta_0\eta_n = +1$ then the
transition amplitude $f_{n}({\bf k},{\bf k}_{0},{\bf R})$ is given
by
\begin{equation} \label{fn_even}
f_{n}({\bf k},{\bf k}_0,{\bf R})=-2\,\Omega_{n}^{(+)}
\cos({\bf k},{\bf R})
\cos({\bf k}_0,{\bf R})-
2\,\Omega_{n}^{(-)}
\sin({\bf k},{\bf R})
\sin({\bf k}_0,{\bf R}), \hspace{6mm}
\end{equation}
otherwise if $\eta_0\eta_n = -1$ then the amplitude
$f_{n}({\bf k},{\bf k}_0,{\bf R})$ have another form
\begin{equation} \label{fn_odd}
f_{n}({\bf k},{\bf k}_0,{\bf R})=2i\,\Omega_{n}^{(-)}
\cos({\bf k},{\bf R}) \sin({\bf k}_0,{\bf R}) -
2i\,\Omega_{n}^{(+)}
\sin({\bf k},{\bf R}) \cos({\bf k}_0,{\bf R}),
\end{equation}
where factors $\Omega_{n}^{(\pm)} = (\Lambda^{(\pm)-1})_{n0}$. The
first item of the Eqs. $(\ref{fn_even})$ and $(\ref{fn_odd})$ is
even function (under reflection ${\bf k} \rightarrow -{\bf k}$) that
conform to the $\Sigma^{+}_{g}$ wave and the second item is odd function
that corresponds $\Sigma^{+}_{u}$ wave.

\paragraph{\hspace{5mm}
\it The example of diatomic, homonuclear molecules in the
2-$\Sigma$-state approximation:} it's a special case of the
preceding model and so we describe the case briefly, introducing
additional useful notations. The matrices $A_{1}$ and $A_{2}$ are
given
\begin{equation}\label{MP}
A_{1}= \left(
\begin{array}{cc}
\alpha_{0} & \lambda \\
\lambda    & \alpha_{1}
\end{array} \right), \hspace{10mm}
A_{2}= \left(
\begin{array}{cc}
\alpha_{0}    & \eta_{0}\eta_{1} \lambda \\
\eta_{0}\eta_{1}  \lambda & \alpha_{1}
\end{array} \right),
\end{equation}
where $\lambda$ is the real coupling channel parameter.
The equations both for amplitudes $f_{n}({\bf k},{\bf k}_0,{\bf R})$
(see Eqs. $(\ref{fn_even})$, $(\ref{fn_odd})$) and for factors
$\Omega^{(\pm)}_n$ remain valid in this case.
In particular the factors $\Omega_{n}^{(\pm)}$ have the form
\begin{equation} \label{Omegan}
\left(
\begin{array}{l}
\Omega^{(\pm)}_{0} \\
\Omega^{(\pm)}_{1}
\end{array}
\right)=
\frac{1}{\theta^{(\pm)}_{0}\theta^{(\pm)}_{1}- \lambda^{2 \mathstrut}}
\left(
\begin{array}{c}
\theta^{(\pm)}_{1} \\
{\scriptstyle -} \lambda
\end{array}
\right).
\end{equation}

\section{The adiabatic-nuclei approximation}

Further we suppose that adiabatic-nuclei approximation is valid
for both rotations and vibrations account. Initially this
approximation was applied by Drozdov $\cite{Dr1955}$, Chase
$\cite{Ch1956}$ and Oksyuk $\cite{Ok1965}$.
The adiabatic approximation in ZRP model
was developed by Demkov and Ostrovsky $\cite{DO1975}$ and Drukarev
and Yurova $\cite{DYu1977}$ (see also Ostrovsky and
Ustimov $\cite{OU1981}$). Differential cross sections of the
electron-rotational-vibrational transitions can be expressed via
corresponding matrix elements of the electron transition amplitude
which obtained in space-fixed nuclei approximation by the formula
$$
\frac{d\sigma}{d\Omega}(nv'j'm' \leftarrow 0vjm)=\frac{k_{n}}{k_{0}}\,
|\, \langle nv'j'm' |f_{n}({\bf k},{\bf k}_0,{\bf R})|0vjm \, \rangle\, |^{2},
$$
where $(0vjm)$ - initial quantum numbers and $(nv'j'm')$ - final quantum numbers of
the electron-rotational-vibrational molecular states
$$
\langle {\bf R}| nvjm \, \rangle = R^{-1}\, X_{n v}(R)\,
Y_{j}^{m}(\widehat{\bf R\!}\,).
$$
Here $Y_{j}^{m}(\widehat{\bf R\!}\,)$ are spherical harmonics
$\cite{VMK1975}$ and $X_{nv}(R)$ are vibrational
harmonics (see Appendix). Further, we omit '$\wedge$' in arguments
if it does not lead to misunderstanding. In
general, the vibrational molecule energies essentially exceed the
rotational energies thereby we can neglect the rotational
contributions to the vibrational harmonics. The outgoing electron
momentum $k_n=|{\bf k}|$ and incoming electron momentum $k_0=|{\bf
k}_0|$ are directly related by law of conservation of energy (we
use atomic units throughout)
$$
\frac{k_{n}^2}{2}+E_{nv'}=\frac{k_{0}^2}{2} + E_{0v},
$$
where $E_{nv'}$ - electron-vibrational state energy,
some function of the arguments $n, v'$.

Since the rotational energy levels of a molecule are so closely
spaced, it is often impossible to resolve particular final states
in a scattering experiment. In this case observed differential
cross sections are averaged over initial rotational states $(j,m)$
and hence it is summed over the final rotational states $(j',m')$
$$
\frac{d\sigma}{d\Omega} (nv' \leftarrow 0v)=
\sum_{j=0}^{\infty}\, \frac{b_{j}}{2j+1}\, \sum_{m=-j}^{j}
\ \, \sum_{j'=0}^{\infty} \, \sum_{\ \ m'=-j'}^{j'}
\, \frac{d\sigma}{d\Omega}(nv'l'm' \leftarrow 0vlm),
$$
$b_{j}$ are relative populations of the rotational states.
The summation over initial and final rotational molecular states can
be exactly realized by using some simple properties of spherical harmonics.
Final formula for observed differential cross sections have the form
\begin{equation} \label{DCS}
\frac{d\sigma}{d\Omega}(nv' \leftarrow 0v) =
\frac{1}{4\pi}\, \frac{k_{n}}{k_{0}} \int d\widehat{\bf R\!}\
\left|\, \int_{0}^{\infty} \overline{X_{n v'}(R)}
f_{n}({\bf k},{\bf k}_{0},{\bf R}) X_{0 v}(R)
\, d R \ \right|^2.
\end{equation}
Differential cross section for pure electronic transition can be
obtained by summation over complete set of final vibrational
substates (generally - including continuous spectrum)
\begin{equation} \label{pDCS}
\frac{d\sigma}{d\Omega}(n \leftarrow 0v) =\frac{1}{4\pi}\, \frac{k_n}{k_0}
\int_{0}^{\infty} dR\ |X_{0 v}(R)|^2\,
\int d\widehat{\bf R\!}\
|f_{n}({\bf k},{\bf k}_{0},{\bf R})|^2.
\end{equation}
Rigorously speaking, the formula is certainly correct only if
the electron-vibration state energy $E_{nv'}$ and, therefore,
outgoing electron momentum $k_n$ are independent on vibrational
quantum number $v'$.

Integral cross section of the electronic-vibrational and pure
vibrational transitions can be evaluated by integration over all
angles $\widehat{\bf k}$
\begin{equation} \label{ICS}
\sigma(nv' \leftarrow 0v) = \int d\widehat{\bf k}\ \,
\frac{d\sigma}{d\Omega}(nv' \leftarrow 0v).
\end{equation}

\paragraph{\hspace{5mm}
\it Differential cross sections of pure electronic transitions:} Let
us start from a  differential cross section for pure electron
transitions. Taking into account the analytical form Eqs.
$(\ref{fn_even})$ and $(\ref{fn_odd})$ we can perform exact
averaging over all molecular orientations (see Eq.
$(\ref{pDCS})$). Final result is given by
$$
\frac{d\sigma}{d\Omega}(n \leftarrow 0v) = \frac{k_n}{k_0}
\int_{0}^{\infty} dR\ |X_{0 v}(R)|^2 \cdot
$$
\begin{equation} \label{pureDCS}
\left[ \left|\Omega^{(+)}_{n}\right|^2 {\cal A}^{(+)}({\bf k},{\bf
k}_0) + \left|\Omega^{(-)}_{n}\right|^2 {\cal A}^{(-)}({\bf
k},{\bf k}_0) + 2 \Re \left( \Omega^{(+)}_n
\overline{\Omega_n^{(-)}\!} \right) {\cal A}({\bf k},{\bf k}_0)
\right],
\end{equation}
here
$$
{\cal A}^{(\pm)}({\bf k},{\bf k}_0)= 1 \pm \frac{\sin(2k_0R)}{2k_0R}
\pm \eta_0 \eta_n
\frac{\sin(2kR)}{2kR} + \eta_0\eta_n
\left( \frac{\sin(2|{\bf k}{\scriptstyle -}{\bf k}_0|R)} {4|{\bf
k}{\scriptstyle -}{\bf k}_0|R} + \frac{\sin(2|{\bf k}{\scriptstyle
+}{\bf k}_0|R)} {4|{\bf k}{\scriptstyle +}{\bf k}_0|R} \right),
$$
and
$$
{\cal A}({\bf k},{\bf k}_0)=\eta_0 \eta_n \left( \frac{\sin(2|{\bf
k}{\scriptstyle -}{\bf k}_0|R)} {4|{\bf k}{\scriptstyle -}{\bf
k}_0|R} - \frac{\sin(2|{\bf k}{\scriptstyle +}{\bf k}_0|R)}
{4|{\bf k}{\scriptstyle +}{\bf k}_0|R} \right)
$$
This result has a specific features therefore we put in the
separate paragraph.

% Taking into account the analytical form
%Eqs. $(\ref{fn_even})$ and $(\ref{fn_odd})$ one can perform exact
%averaging over all molecular orientations.

\paragraph{\hspace{5mm}
\it Differential cross sections of electronic-vibrational
transitions:} now we consider differential cross sections of
electron-vibrational transitions. Using the analytical form (see
Eqs. $(\ref{fn_even})$ and $(\ref{fn_odd})$) of electron
transition amplitude $f_{n}({\bf k},{\bf k}_0,{\bf R})$ we can
perform exact averaging over all directions of $\widehat{\bf
R\!}\, $. In order to make it we begin with  electron transition
amplitude transforming into infinite sum of the spherical
harmonics with unit vector $\widehat{\bf R\!}\, $ as the argument
\begin{equation}\label{tf_n}
\begin{array}{r}
\displaystyle f_{n}({\bf k},{\bf k}_0,{\bf R})=-4\pi \sum_{l} \sum_{m=-l}^{l}
i^l\, Y_{l}^{m}({\bf R})
\left[ \left(\Omega_{n}^{(+)}-\Omega_{n}^{(-)}\right)\,
j_{l}(|{\bf k} {\scriptstyle +} {\bf k}_0|R)\,
\overline{Y_{l}^{m}({\bf k} {\scriptstyle +} {\bf k}_0)} \, +\, \right. \\
\left. \left(\Omega_{n}^{(+)}+\Omega_{n}^{(-)}\right)\,
j_{l}(|{\bf k} {\scriptstyle -} {\bf k}_0|R)\,
\overline{Y_{l}^{m}({\bf k} {\scriptstyle -} {\bf k}_0)} \, \right]. \,
\end{array}
\end{equation}
where the value $l$ is even in case of $\eta_0 \eta_n =+1$ and odd
if $\eta_0 \eta_n =-1$,
$j_{l}(z)=\sqrt{\frac{\displaystyle \pi} {\displaystyle 2z
\mathstrut}} J_{l+1/2}(z)$, and $J_{\,l + 1/2}(z)$ are the
Bessel functions with indices $l + 1/2$. Plugging the transformed
amplitude $(\ref{tf_n})$ into the expression for the differential
cross section (see Eq. $(\ref{DCS}))$ and averaging over angular
variables $\widehat{\bf R \!}\,$  using the addition theorem for
spherical harmonics yields
\begin{equation} \label{TMP}
\sum_{m=-l}^{l} |Y_{\, l}^{m}({\bf x})|^2= \frac{2
l+1}{4\pi},\hspace{7mm}\mbox{and}\hspace{5mm} \sum_{m=-l}^{l}
\overline{Y_{\, l}^{m}({\bf x})}\, Y_{\, l}^{m}({\bf
y})=\frac{2l+1}{4\pi}\, P_{l}(({\bf x},{\bf y})).
\end{equation}
Here the functions $P_{l}(z)$ are Legendre polynomials $\cite{VMK1975}$.
Finally we obtain the expression for differential cross sections of
the electron-vibrational excitations
\begin{equation} \label{av_DCS}
\frac{d\sigma}{d\Omega}(nv' \leftarrow 0v)=
\frac{k_n}{k_0}\, \sum_{l}\, (2 l + 1)
\left\{\left|g_{+}^{(l)}\right|^2 + \left|g_{-}^{(l)}\right|^2 + 2 \Re
\left(g_{+}^{(l)}\overline{g_{-}^{(l)}}\right) P_{l}(\cos \vartheta) \right\},
\end{equation}
where we use following notation
$$
g_{\pm}^{(l)}=\int_{0}^{\infty}
j_{l}(|{\bf k} {\scriptstyle \pm} {\bf k}_0|R)
\left(\Omega_{n}^{(+)} \mp \Omega_{n}^{(-)} \right)\,
\overline{X_{n v'}(R)} \,
X_{0 v}(R) \, d R,
$$
$$
\cos \vartheta=\frac{({\bf k}{\scriptstyle +}{\bf k}_{0},\
{\bf k}{\scriptstyle -}{\bf k}_{0})}
{|{\bf k} {\scriptstyle +} {\bf k}_0|\, |{\bf k}{\scriptstyle -}{\bf k}_0|}.
$$
We omit the indices $n,v',v$ in the notation $g_{\pm}^{(l)}$. To
some extent this expression is the generalization of a Drukarev
and Yurova formula (see Ref. $\cite{DYu1977}$) for pure
vibrational-rotational excitations at fixed ${\bf R}$. However, as
distinct from above mentioned result the formula $(\ref{av_DCS})$
arise from averaging over all initial and summing over final
rotational molecular states so as only electron-vibrational
excitations are  taken into account.

\paragraph{\hspace{5mm}
\it Integral cross section  of a pure electron transition:} There
exist two alternative ways to calculate an integral cross section
(ICS). In the first one we integrate the averaged differential
cross section (see Eq. $(\ref{DCS})$) over all directions of the
outgoing electron momentum ${\bf k}$. Otherwise we can integrate
the differential cross section with some fixed molecular
orientation $\widehat{\bf R\!}\, $ over $\widehat{\bf k}$ and
average over all direction of incoming electron momentum ${\bf
k}_0$. Hereinafter we adhere to the second approach because it is
more simple. The final result has the form
$$
\sigma(n \leftarrow 0v)=4\pi \frac{k_{n}}{k_{0}}
\int_{0}^{\infty} \left\{
\left|\Omega_{n}^{(+)}\right|^2 \left(1+\eta_0\eta_n
\frac{\sin(2kR)}{2kR}\right)
\left(1+\frac{\sin(2k_0R)}{2k_0R}\right)+ \right.
$$
\begin{equation}
\left. \hspace{37mm} \left|\Omega_{n}^{(-)}\right|^2
\left(1-\eta_0\eta_n \frac{\sin(2kR)}{2kR}\right)
\left(1-\frac{\sin(2k_0R)}{2k_0R}\right) \right\}\,
|X_{0v}(R)|^2 dR.
\end{equation}

\paragraph{\hspace{5mm} \it Integral cross sections for
electron-vibration transitions:}
In order to calculate the ICSs for electron-vibration transitions we also
adhere to the second approach, because it allows to achieve one's
purpose without Clebsch-Gordan coefficients utilization.
It's convenient before integration over
$\widehat{\bf k},\widehat{\bf k}_0$ to represent the electron
transition amplitudes $f_{n}({\bf k},{\bf k}_0,{\bf R})$ (see Eqs.
$(\ref{fn_even})$, $(\ref{fn_odd})$) as infinite sum over
spherical harmonics with unit vectors $\widehat{\bf k}$, and
$\widehat{\bf k}_0$ as arguments
$$
f_{n}({\bf k},{\bf k}_0,{\bf R})=32\, \pi^2 \cdot
$$
$$
\left[ -\Omega_{n}^{(+)}\sum_{even\ l}\ \sum_{\ m=-l}^{l}\ \sum_{L}\
\sum_{\ M=-L}^{L}\ i^{(l+L)} \, j_{l}(k_0 R)\, j_{L}(k R)\,
Y_{l}^{m}({\bf R})
\overline{Y_{l}^{m \mathstrut}({\bf k}_0)}
Y_{L}^{M}({\bf R})
\overline{Y_{L}^{M \mathstrut}({\bf k})} \, + \right.
$$
$$
\left. \Omega_{n}^{(-)}\sum_{odd\ l_0}\ \sum_{\ m_0=-l_0}^{l_0}\ \sum_{L_0}\
\sum_{\ M_0=-L_0}^{L_0}\ i^{(l_0+L_0)} \, j_{l_0}(k_0 R)\, j_{L_0}(k R)\,
Y_{l_0}^{m_0}({\bf R})
\overline{Y_{l_0}^{m_0 \mathstrut}({\bf k}_0)}
Y_{L_0}^{M_0}({\bf R})
\overline{Y_{L_0}^{M_0 \mathstrut}({\bf k})} \right],
$$
where summations are performed over even $L$ and odd $L_0$ in case
of $\eta_0\eta_1=+1$ and over odd $L$ and even $L_0$ in case
of $\eta_0\eta_1=-1$. In accordance with foregoing the integral cross
section can be obtained by the following integration
$$
\sigma(nv' \leftarrow 0v) = \frac{1}{4\pi}\, \frac{k_n}{k_0}
\int d\widehat{\bf k} \int d\widehat{\bf k}_0
\left| \int_{0}^{\infty} \overline{X_{n v'}(R) \mathstrut}\,
f_{n}({\bf k},{\bf k}_0,{\bf R}) X_{0v}(R)\, d R \, \right|^2.
$$
Using the spherical harmonic orthogonality and Eq. $(\ref{TMP})$ we obtain for
integral cross section
\begin{equation} \label{ICS_n}
\begin{array}{c}
\displaystyle \sigma(nv' \leftarrow 0v)=16\pi \frac{k_n}{k_0}\, \cdot \\
\displaystyle \left\{ \sum_{even\ l}\,\sum_{L}\, (2l+1)(2L+1)
\, \left|q_{+}^{(lL)} \right|^2\, +\,
\sum_{odd\ l_0}\,\sum_{L_0}\, (2l_0+1)(2L_0+1)
\left|q_{-}^{(l_0 L_0)}\right|^2\, \right\} ,
\end{array}
\end{equation}
here we use the following notation
$$
q_{\pm}^{(lL)}=\int_{0}^{\infty}\,
\Omega_{n}^{(\pm)} j_{l}(k_0 R)\, j_{L}(k_n R)\,
\overline{X_{n v'}(R)}\, X_{0v}(R)\, d R.
$$

\section{Applications and discussion}
\paragraph{\hspace{5mm} \it Molecule $H_2$,
$X\, ^{1}\Sigma_g^{+} \rightarrow a\, ^{3}\Sigma_{g}^{+}$ transition:}
selecting the ZRPs parameters, we proceeded from the assumption that
elements of the scattering length matrix
$$
(A_1)^{-1}=\left(
\begin{array}{cc}
a & c \\
c & b \\
\end{array} \right)
$$
are independent of number of channels $N$. Therefore, the parameter $a$
can be chosen equal $1/0.35$ (see Ref. $\cite{DYu1977}$) and
$c,b$ are adjusting parameters. In our calculation, we also used
the vibration  quantum $\omega_{0}=2\times10^{-2}$, anharmonicity constant
$\ae_{0}=5.74\times10^{-4}$ and equilibrium internuclear distance $2R_0=1.401$.

\begin{figure}
\centering
\includegraphics[scale=0.9]{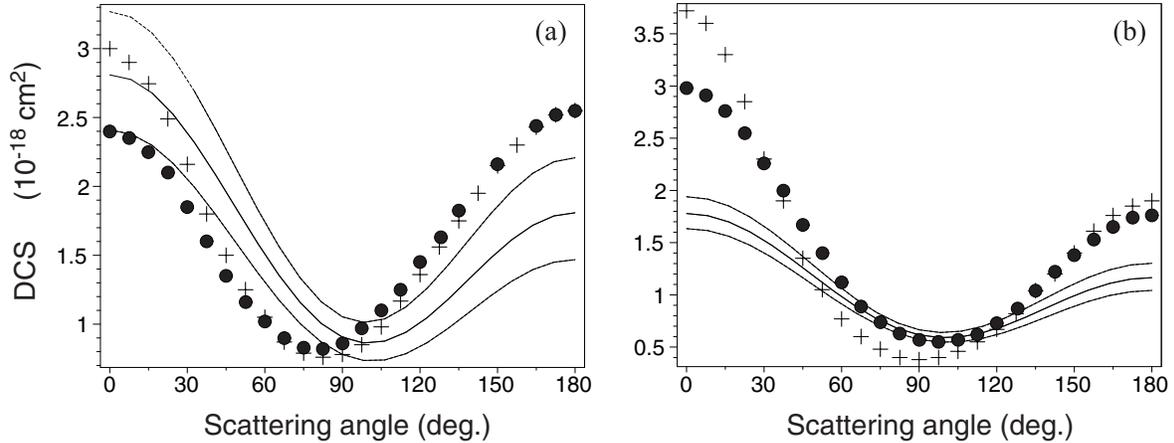}
\parbox[t]{0.92\textwidth}{
\caption{DCS for the pure $X\, ^1\Sigma_g^{+} \rightarrow a\, ^3\Sigma_g^{+}$
transition $\hbox{H}_2$ by electron impact at (a) 15 eV and (b) 18 eV.
Solid lines, presents our calculation for $b=1.35$ (lower line), 1.40 (midline),
and 1.45 (upper line);
full circles, SMC results of Lima,
et al. $\cite{Lim1988}$; crosses, MCF results of Lee, et al. $\cite{Lee1998}$}
\label{f:H2dcs}}
\end{figure}

In Figs. $\ref{f:H2dcs}$, we present our calculated differential
cross sections (DCSs) for pure electron-impact electronic excitation
for $c=0.63$ and $b=1.35$, $1.40$, and $1.45$.
We also compare our DCSs at some selected energies with the Schwinger
multichannel (SMC) results of Lima, et al. $\cite{Lim1988}$
and method of continued fractions (MCF) results of Lee, et al. $\cite{Lee1998}$.
In general, there is good qualitative agreement.
However, our and their results differs in the forward and backward directions,
particularly for impact energies above 17-18 eV.

\begin{figure}
\centering
\includegraphics[scale=0.9]{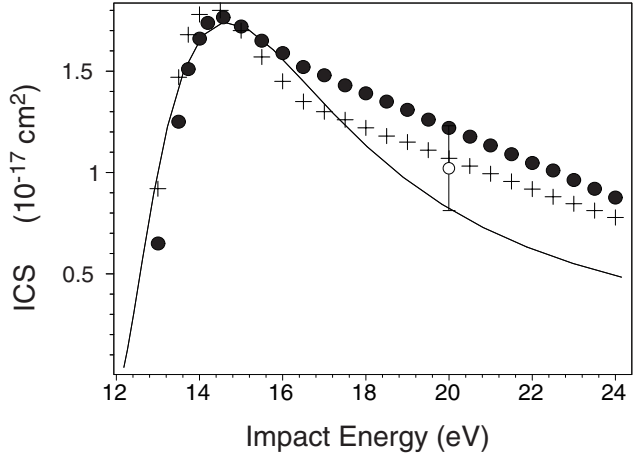}
\parbox[t]{0.9\textwidth}{
\caption{ICS for the pure $X\, ^1\Sigma_g^{+} \rightarrow a\,
^3\Sigma_g^{+}$ electron transition by electron impact in the
12-24 eV range. Solid line, present our calculation; full circles,
SMC results of Lima, et al. $\cite{Lim1988}$; crosses, MCF results
of Lee, et al. An experimental point is also shown.
$\cite{Lee1998}$} \label{f:H2ics}}.
\end{figure}

Fig. $\ref{f:H2ics}$ show our integral cross sections for pure
electron-impact electronic excitation. In our calculation, the parameters
$c=0.63,$ and $b=1.4$ were used. We compare our ICSs with the
SMC results of Lima, et al. $\cite{Lim1988}$ and MCF results of Lee, et al.
$\cite{Lee1998}$. Our calculated ICSs are in good agreement, both
qualitatively and quantitativly, with their theoretical data
in the 13-17 eV range. Our ICSs smaller then their results for
impact energies above 17-18 eV.

We think that possibilities of the ZRP methods allow to improve
the results if

1) to use better approximations for a potential well in which the
molecule oscillates,

2) to incorporate a separable potential into the theory (e.g.,
\cite{Bal,Pr1997}),

3) account other channels of electron excitations.

All these developments of the model could diminish the difference
between the plots obtained by this approach, simulations "ab
initio" and experiments.

\section{Conclusion}
We claim that the next step of our work is to unify both
multichannel and higher partial modes \cite{Bal} descriptions.
This way we evaluate amplitudes of the $N_2$ electron-molecular
scattering at low energies to fit recent experiments \cite{Zub}.
The results will be published elsewhere. The approach has one more
natural generalization for multicenter scattering \cite{DR1970,Szm1999}
\section{Acknowledgements}
We acknowledge consultations and priceless advices of
V. Ostrovsky and I. Yurova and
  discussions with J. Sienkiewicz and M. Zubek.

\appendix
\section{The vibrational harmonics $X_{n v}(R)$} \label{APP_VH}
In this work we approximate the theoretical energy "curves" of
molecular states by the Morse potentials
$$
U_{n}(R)=\frac{\omega_{n}^2}{4\ae_{n}}
\left[ 1- \exp\left(\displaystyle -2\sqrt{2\mu\ae_{n}}(R-R_{n})\right) \right]^2+
U_{n},
$$
where $\mu$ is reduced mass of molecule, $\omega_{n}, \ae_{n}$ -
corresponding vibration  quantum and anharmonicity constant of the electronic
state $n$, $U_{n}$ are energies at equilibrium internuclear
distances $2R_{n}$, and $2R$ - internuclear distance.
The choice arbitrariness of the parameters $U_{n}$ can be restricted
by ground state energy fixation. For that we demand ground state energy
be equal zero. In this case
$$
U_{0}=-\frac{\displaystyle \omega_0}{\displaystyle 2}+
\frac{\displaystyle \ae_0}{\displaystyle 4}.
$$
The vibration harmonic
$X_{n v}(R)$ satisfied the radial Schr\"{o}dinger equation
$$
-\frac{1}{8\mu}X''_{n v} +U_{n}(R) X_{n v} =
E_{n v} X_{n v},
$$
here argument $R$ is omitted. The physically reasonable solution of this equation
can be expressed via orthogonal
Laguerre polynomials $L_{v}^{\xi}(z)$
$$
X_{n v}(R)= (C_{n v})^{-1/2}\, z^{\xi/2}\, \exp(-z/2)\, L_{v}^{\xi}(z),
$$
where we use the following notation
$$
z=\frac{\omega_{n}}{\ae_{n}}\, \exp(-2\sqrt{2\mu\ae_{n}}(R-R_{n})),
$$
$$
\hspace{10mm} \xi=\frac{\omega_{n}}{\ae_{n}}-2v-1.
$$
The normalization factors $C_{n v}$ are defined by the normality condition. If
functions $X_{n v}(R)$ are normalized to unity then the high accurate approximations
for normalization factors are given
$$
C_{n v}=\frac{\Gamma(\xi+v+1)}{\displaystyle v!\, \xi \sqrt{8m\ae_{n} \mathstrut}}.
$$
The energy levels $E_{n v}$ constitute the finite sequence
$$
E_{n v} = \omega_{n} \left(v+\frac{1}{2}\right) - \ae_{n} \left( v
+ \frac{1}{2} \right)^2 + U_{n},\hspace{15mm}
v=0,1,..,v_{m}.
$$

\end{document}